\documentclass[reprint,twocolumn,showpacs,showkeys,letter]{revtex4-1}

\usepackage{latexsym,graphicx,graphics}
\usepackage{appendix}
\usepackage{amsmath,amssymb,epsfig}
\usepackage[utf8]{inputenc}
\usepackage[mathscr]{euscript}
\usepackage{float}

\newcommand{\abs}[1]{\left| #1 \right|}

\newcommand{\bb}[1]{ \mbox{\boldmath$ #1$}}

\newcommand{\rv}{\bb r}

\newcommand{\sumex}[3] {\sum_{#1}^{#2}\! ^{#3}}

\newcommand{\Eq}[1]{Eq.~(\ref{#1})}
\newcommand{\Eqs}[2]{Eqs.~(\ref{#1})--(\ref{#2})}

\newcommand{\unit}[1]{\bb{\hat{#1}}}

\begin{document}

\title{Planar One-way Guiding in Periodic Particle Arrays with Asymmetric Unit Cell and General  Group-Symmetry Considerations}

\author{Y. Mazor}
\email{yardenm2@mail.tau.ac.il}

\author{Y. Hadad}%
\email{yakhadad@gmail.com}
\altaffiliation[Currently at ]{Department of Electrical Engineering, University of Texas at Austin}

\author{Ben Z. Steinberg}
\email{steinber@eng.tau.ac.il}
\affiliation{
School of Electrical Engineering, Tel-Aviv University, Ramat Aviv, Tel-Aviv 69978 Israel
}%

\date{\today}

\begin{abstract}
We develop a general theory for one-way optical guiding in magnetized periodic particle arrays. Necessary conditions for a non-even dispersion curves are derived and presented in the context of Frieze symmetry-groups. It is shown, for example, that one-way guiding can be supported in particle \emph{strips} consisting of geometrically isotropic particles arranged in transversely asymmetric arrays. Specific examples consist e.g.~two parallel isotropic particle chains with different periods.
The previously studied one-way effect based on the two-type rotation principle is shown to be a special case. In the latter the exclusion of the appropriate Frieze-symmetries is achieved in a single linear chain by associating a geometric rotation to each particle, thus providing the narrowest possible one-way waveguides. It is also shown that nearly any randomly created period may result in uneven dispersion and one-way guiding.
\end{abstract}

\maketitle

\section{Introduction}

One-way waveguides are important building blocks for many functional systems in acoustics and  electromagnetics.
They find applications as isolators and circulators, as a mean to reduce disorder effects in  waveguides \cite{YUFAN}, or even as a mean to match waveguide to a load \cite{Engheta_MatchedLoad} or to a leaky wave nano-antenna \cite{HadadSteinberg_Antenna}.

Non-reciprocity and one-way guiding can be achieved either by medium's non-linearities, by spatiotemporal modulation, or by magnetic biasing.
Non-linear non-reciprocal structures \cite{Wabnitz_1986,GalloFejer_2001,SolacicFan_2003,Trendafilov_2010,FanQi_2012} use the fact that reciprocity does not imply symmetric field distributions of the forward and backward propagating modes. Therefore, spatially-varying material nonlinearity created in an initially reciprocal waveguide, may affect the two modes differently and eventually result in non-reciprocity. Since it is based on nonlinear effects, this method requires large volumes, it is power-dependent, and may lead to signal distortion.
Alternatively, spatiotemporal variation of the medium's properties may violate reciprocity.  For example, by using  a traveling wave to modulate the medium's constitutive parameters one may asymmetrically add a ``momentum'' (wave number) bias to the otherwise-symmetric mode picture.  Several non-reciprocal components and one-way waveguides have been proposed using this idea, for electromagnetic waves \cite{YuFan_2009,LiraLipson_2012} as well as for acoustic waves \cite{ZanjaniLukes_2013}.
These structures are linear and magnetic-biasing free. However, the modulation effect is usually quite weak thus yielding a narrow one-way bandwidth and a large device, many wavelengths long.
The third way to violate reciprocity is using magneto-optical effects produced by static magnetic biasing under which ferrites and plasmas become gyrotropic. Then, if a waveguide is asymmetrically loaded, isolation can be obtained \cite{Collin}. Many of the recently proposed one-way schemes in optics are based on this idea \cite{YUFAN,FigotinVitebsky1,FigotinVitebsky2,Jung,YUFAN2,Haldane,BiRoss_2011, KhanikaevKivshar_2010}. However, these structures have lateral width of at least one wavelength.
Ring resonators have been proposed as a mean to enhance non-reciprocity and reduce the device's size. The  degeneracy of a pair of counter propagating modes can be removed   by azimuthal spatiotemporal modulation \cite{SounasAlu_2013,SounasAlu_2014,FleuryAlu_2014}, or by magnetization \cite{BiRoss_2011} -  yielding a mode-Q-factor dependent isolation with device footprint of the order of a wavelength.

Along an apparently different research endeavor,
linear periodic chains of equally-spaced identical plasmonic nano-particles were studied in a number of publications \cite{Gerardy}-\cite{Markel_CurvedChains}. Particle chains were proposed as guiding structures and junctions in \cite{Quinten}-\cite{Capolino_Chain}, as surface waves couplers \cite{Lomakin}, and as polarization-sensitive waveguides \cite{Lomakin2}. When the inter-particle distance is sufficiently small, the guided optical wave in such chains is highly localized, i.e.~it's spatial width is much smaller than the wavelength that corresponds to the operation frequency. Hence, they are termed as Sub-Diffraction Chains (SDC's). The chain Green's function was calculated in \cite{Markel_2}, and
the SDC's modal features were studied using a general approach and spectral analysis in \cite{EnghetaChain, Capolino_Chain} and were also considered in \cite{Markel_3}. Green's function theories revealing all the wave constituents that can be excited in these structures, were developed and discussed in detail in \cite{HadadSteinbergPRB}. Scattering due to structural disorder and its effect on the modes were studied in \cite{Markel_2}, and in \cite{AluPaper}. Propagation in curved SDC's has been considered recently in \cite{Markel_CurvedChains}. The effect of liquid-crystal host of SDC's has been studied in \cite{Pike_Stroud_SDC_LC}, where the interesting option of dynamically controlled SDC is suggested.

In \cite{HadadSteinberg_SpiralChain,MazorSteinberg_PlanarOneWay,HadadMazorSteinberg_GreensOneWay} a family of magnetized plasmonic SDC's one-way waveguides, based on the `two-type rotation'' principle was proposed for one-way guiding and optical isolation. As has been shown there, exposing a conventional spherical particles SDC to external magnetic bias may formally create non-reciprocity due to the Faraday rotation induced in the plasmonic particles. However, this is \emph{not sufficient for optical isolation} since the resulting dispersion $\omega(\beta)$ is still an even function and no preferred direction is created.
To achieve the one-way property, the suggested structures consist of magnetized, \emph{non-spherical}, plasmonic particles arranged as a linear array. Each particle in the array is rotated with respect to its neighbors, rendering the chain chiral. The interplay between the field polarization rotation due to the gyrotropy of the resonant particles and the structural chirality is utilized in order to enhance non-reciprocity and create uneven dispersion. Significant optical isolation and one-way guiding are then obtained at operation points near the light-cone - see \cite{HadadSteinberg_SpiralChain,MazorSteinberg_PlanarOneWay,HadadMazorSteinberg_GreensOneWay}.
These structures have two main advantages over other one-way schemes. First, as said above, the effective lateral width of the guided mode is much smaller than the wavelength. Second, it has been shown that the required magnetic bias is considerably weaker than the bias used in other magnetization-based one-way schemes. This bias may be further reduced by utilizing the enhanced Faraday rotation in a plasmonic particle consisting of a core-shell geometry such as the particles studied in \cite{Graphene_Magnetized,CoreShell_Faraday_Engheta}.
However, despite their appealing properties, a nanoscale fabrication of the proposed structures might be somewhat challenging due to the apparent requirement for a precisely characterized geometrical features in the particles.

\begin{figure}
        \includegraphics[width=6cm]{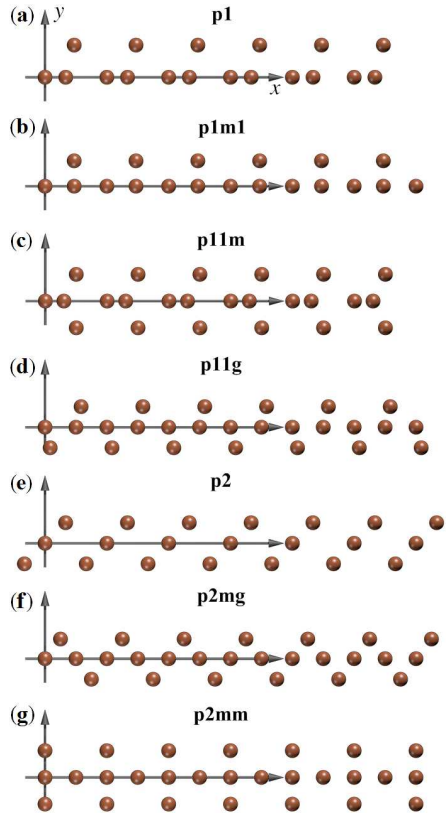}
    \caption{A periodic particle strip belongs to at least one of the seven Frieze symmetry groups, invariant under the following operations (in addition to invariance under $\mathcal{T}_D$): (a) No other symmetry - the {\bf p1} group. (b) $\mathcal{R}_v$ - the {\bf p1m1} group. (c) $\mathcal{R}_h$ - the {\bf p11m} group. (d) Glide reflection: $\mathcal{R}_h\mathcal{T}_d,\,\, d<D$ - the {\bf p11g} group. (e) $\mathcal{R}_\pi$ - the {\bf p2} group. (f) $\mathcal{R}_v$, $\mathcal{R}_\pi$, and $\mathcal{R}_h\mathcal{T}_d,\,\, d=D/2$ - the {\bf p2mg} group. (g) $\mathcal{R}_v$, $\mathcal{R}_h$, and $\mathcal{R}_\pi$ - the {\bf p2mm} group.
    }
    \label{fig1}
\end{figure}

Here we approach the problem of enhanced non-reciprocity from a wider point of view. Our study provides the \emph{necessary} conditions for uneven dispersion curves and one-way guiding in planar particle arrays, some examples of which are shown schematically in Fig.~\ref{fig1}. The particles may or may not be identical and symmetric.  The necessary conditions for uneven dispersion are characterized in the Frieze symmetry-groups framework. The previously reported one-way chains based on the two-type rotation principle \cite{HadadSteinberg_SpiralChain,MazorSteinberg_PlanarOneWay,HadadMazorSteinberg_GreensOneWay} are shown to be a special case that conveys the narrowest structure (but not the simplest to fabricate). One of the important conclusions of the present study is that \emph{almost any random construction} of the structure's period would result, under magnetic bias, in uneven dispersion and may function as an optical isolator. In fact, due to the structural asymmetry requirements for uneven dispersion studied here, and due to the fact that fabrication errors are unavoidable, it may turn out that under magnetic bias uneven dispersion is easier to obtain than even dispersion.

\section{Formulation}

 We refer to the periodic particle ``strips'' shown schematically in Fig.~\ref{fig1}. For convenience, let us denote the plane on which the strips lie as the $(x,y)$ plane, $x$ being the strip axis. The strip period is $D$. To study their symmetry properties we define the following translation by $d$, reflection about horizontal/vertical line, and rotation by $\pi$ operators in the $(x,y)$ plane
\begin{subequations}\label{eqs1}
\begin{eqnarray}
\mathcal{T}_d(x,y) & \mapsto & (x+d,y)\label{Eq1a}\\
\mathcal{R}_h(x,y) & \mapsto & (x,-y) \label{Eq1b}\\
\mathcal{R}_v(x,y) & \mapsto & (-x,y) \label{Eq1c}\\
\mathcal{R}_\pi(x,y) & \mapsto & (-x,-y).\label{Eq1d}
\end{eqnarray}
\end{subequations}
We emphasize that the symmetry of any of the structures presented, e.g. in Fig.\ref{fig1} emerges not only from the \emph{locations} of the particles, but also by the specific properties of the particles populating these locations; e.g.~their \emph{polarizabilities}.
Our aim is to study the electrodynamic properties of the arrays schematized in Fig.~\ref{fig1} and their potential use for enhanced breach of time-reversal symmetries. In particular, we look for the necessary conditions under which they support guided modes with uneven dispersion curves and, eventually, one-way guiding.

The particles are much smaller than the free-space wavelength $\lambda$ at the structure's operation frequency $\omega$. Hence, we use polarizability theory and the discrete-dipole approximation. The structure's period consists of $N$ particles, not necessarily identical, each characterized by its own polarizability matrix $\bb{\alpha}_n,\, n=1,\ldots ,N$.
In the present work, a necessary condition for the asymmetries of $\bb{\alpha}_n$ is achieved by applying a bias magnetic field $\bb{B}_0=\unit{z}B_0$, that affects essentially the particles dipole response in the $x,y$ plane. Therefore the dipoles $\unit{z}$ components are ignored, rendering $\bb{\alpha}_n$ a set of $N$ matrices of $2\times 2$ elements.
We denote the location of the $n$-th particles in the $m$-th period by $\rv_{m,n}$, and the dipole moment excited in this particle by $\bb{p}_{m,n}$. The structure's electrodynamics is governed by the infinite difference equation
\begin{equation}\label{eq2}
\bb{p}_{m,n}=\bb{\alpha}_n\sum_k\sumex{\ell=1}{N}{(m,n)}\, {\bf G}(\rv_{m,n},\rv_{k,\ell})
\bb{p}_{k,\ell}
\end{equation}
where ${\bf G}(\rv,\rv')$ is the free-space Green's function matrix, by which ${\bf G}(\rv,\rv')\bb{p}$ gives the electric field at $\rv$ due to an electric dipole $\bb{p}$ at $\rv'$. Here and henceforth, the absence of summation limits indicates a summation over all integers, and $\sum \! ^{v}$ indicates a summation that excludes the $v$ term.
Due to periodicity, we have
\begin{equation}\label{eq3}
\bb{p}_{m,n}=\bb{p}_{0,n}e^{im\beta D}
\end{equation}
hence \Eq{eq2} can be re-written as the following $2N\times 2N$ matrix equation ($n=1,\ldots,N$)
\begin{subequations}\label{eqs4}
\begin{eqnarray}\label{eq4a}
\!\!\!  &{\bf M}& (\beta){\bf P} \equiv\\
&\equiv &\left(\bb{\alpha}_n^{-1}-{\bf M}_n^D\right)\,\bb{p}_{0,n}-\sumex{\ell=1}{N}{n}\, {\bf M}_{n,\ell}\, \bb{p}_{0,\ell}= \bb{0}\nonumber
\end{eqnarray}
where ${\bf M}_n^D$ are $N$ matrices of $2\times 2$.
The matrices $\left(\bb{\alpha}_n^{-1}-{\bf M}_n^D\right)$ reside in the $N$ diagonal blocks of $2\times 2$ elements each.
${\bf M}_{n,\ell}$ are $2\times 2$ matrices residing in the off-diagonal blocks. ${\bf M}_n^D$ and ${\bf M}_{n,\ell}$ are given by the matrix-sums
\begin{eqnarray}
{\bf M}_n^D(\beta) &=& \sumex{m}{}{0} \,\,{\bf G}(\rv_{0,n},\rv_{m,n})\, e^{im\beta D} \label{eq4b}\\
{\bf M}_{n,\ell}(\beta) &=& \sum_m {\bf G}(\rv_{0,n},\rv_{m,\ell})\, e^{im\beta D}.\label{eq4c}
\end{eqnarray}
\end{subequations}
The structure dispersion is the set of $\beta(\omega)$'s that nullifies the determinant of \Eq{eq4a}. An even dispersion curve is obtained if this determinant is invariant (up to a $\beta$-independent multiplication constant) under the map $\beta\mapsto-\beta$. Below we study the symmetries of \Eq{eq4a} under this map.

First, we note that the matrices ${\bf M}_n^D$ and ${\bf M}_{n,\ell}$ are particle-independent, and their symmetries emerge only from the symmetries of ${\bf G}$ and from the symmetries (if any) of the lattice sites arrangements inside the strip period. These are studied in appendix \ref{AppSymm}. For the diagonal blocks we have,
\begin{equation}\label{eq5}
{\bf M}_n^D(-\beta)={\bf M}_n^D(\beta)=\left[ {\bf M}_n^D(\beta)\right]^T
\end{equation}
and for the off-diagonal blocks
\begin{equation}\label{eq6}
{\bf M}_{n,\ell}(-\beta)={\bf M}_{\ell,n}(\beta)=\left[{\bf M}_{\ell,n}(\beta)\right]^T.
\end{equation}
These properties hold just by a mere periodicity, and are proved in appendix \ref{AppSymm}. Hence, under the mapping $\beta\mapsto -\beta$ \Eq{eq4a} that is written for $\beta$, is mapped to
\begin{eqnarray}\label{eq7}
\!\!\!  &{\bf M}& (-\beta){\bf P} \equiv\\
&\equiv & \left(\bb{\alpha}_n^{-1}-{\bf M}_n^D\right)\,\bb{p}_{0,n}-\sumex{\ell=1}{N}{n}\, {\bf M}_{\ell,n}\, \bb{p}_{0,\ell}=\bb{0}.\nonumber
\end{eqnarray}
Hence, the mapping $\beta\mapsto -\beta$ is manifested by a block-transpose operation on the matrix that consists of the blocks ${\bf M}_{n,\ell}$. Generally, even dispersion curve $\beta(\omega)$ is obtained if there is, up to a multiplication constant, a determinant-preserving transformation by which one can obtain \Eq{eq7} from \Eq{eq4a}. Below we explore the conditions for even dispersion. The simultaneous violation of all these conditions is a necessary condition for uneven dispersion.

\subsection{Symmetric $\bb{\alpha}_n$'s}

If the $\bb{\alpha}_n$'s are symmetric
\begin{equation}\label{eq8}
\bb{\alpha}_n=\bb{\alpha}_n^T,\quad \forall\, n
\end{equation}
then, from \Eq{eq6} we have $\left[{\bf M}(\beta)\right]^T={\bf M}(-\beta)$. Hence $\abs{{\bf M}(\beta)}=\abs{{\bf M}(-\beta)}$ (transpose preserves the determinant), leading to even dispersion.

\subsection{Non-symmetric and different $\bb{\alpha}_n$'s}

Assume now that $\bb{\alpha}_n$ are non-symmetric and also not necessarily identical. For simplicity, let us number the polarizabilities within a period along the positive $\unit{x}$ and $\unit{y}$ directions; i.e.~from left to right and from bottom to top. With the use of the inversion matrix ${\bf I^i}_{2N}$ defined in Appendix \ref{AppInv} \Eq{eqAppInv2}, we have
\begin{subequations}
\begin{equation}\label{eq9a}
{\bf M}(\beta)={\bf I^i}_{2N}{\bf M}(-\beta){\bf I^i}_{2N}
\end{equation}
if the following conditions are satisfied:
\begin{equation}\label{eq9b}
\bb{\alpha}_{N+1-n}=\bb{\alpha}_n, \,\,{\bf M}^D_{N+1-n}={\bf M}^D_n,  \quad \forall\, n
\end{equation}
and
\begin{equation}\label{eq9c}
{\bf M}_{\ell,n}={\bf M}_{N+1-n, N+1-\ell},\qquad \forall\,\ell,n.
\end{equation}
\end{subequations}
These conditions imply inversion symmetry (invariance under $\mathcal{R}_\pi$), that applies on both the particle coordinates [see appendix\ref{AppSymm}, \Eqs{eqAppSymm8}{eqAppSymm9b}], as well as on the particle properties. Since ${\bf I^i}_{2N}$ is determinant preserving, it follows that in inversion symmetric chain structures the dispersion curves are always even, independently of the asymmetric form of each of the individual polarizabilities.

To find another condition for even dispersion, we refer to \Eq{eqAppG3b}, by which applying
  $\mathcal{R}_h$ on the chain is equivalent to matrix transformation by ${\bf I^h}_2$. Consequently, from \Eqs{eq4b}{eq4c} it follows that applying $\mathcal{R}_h$ on the structure is equivalent to the matrix transformations
\begin{subequations}
\begin{eqnarray}
{\bf M}_n^D      &\stackrel{\mathcal{R}_h}{\mapsto}& {\bf I^h}_{2}\,{\bf M}_n^D\, {\bf I^h}_{2}\label{eq10a}\\
{\bf M}_{\ell,n} &\stackrel{\mathcal{R}_h}{\mapsto}& {\bf I^h}_{2}\, {\bf M}_{\ell,n}\, {\bf I^h}_{2}\label{eq10b}
\end{eqnarray}
\end{subequations}
this operation merely changes the signs of the two off-diagonal terms in the matrices ${\bf M}_n^D$ and ${\bf M}_{\ell,n}$. We refer now to \Eqs{eqAppSymm11a}{eqAppSymm11b} that hold if the set of points in a period possesses horizontal reflection symmetry. Hence, using the horizontal inversion matrix operator ${\bf I^h}_{2N}$ defined in appendix \ref{AppInv}, we have
\begin{subequations}
\begin{equation}\label{eq11a}
{\bf M}^T(\beta)={\bf I^h}_{2N}{\bf M}(-\beta){\bf I^h}_{2N}
\end{equation}
if the structure possesses horizontal inversion symmetry and if
\begin{equation}\label{eq11b}
\bb{\alpha}^T_n={\bf I^h}_{2}\bb{\alpha}_n{\bf I^h}_{2},\quad\forall\, n=1,\ldots N.
\end{equation}
\end{subequations}
Since the operations in \Eq{eq11a} are determinant preserving, it follows that under these conditions the dispersion curve is even. Note that the condition in \Eq{eq11b} applies not only to conventional spherical particles, but also to plasmonic particles under magnetization.

Finally, we note that the glide-reflection transformation is defined as $\mathcal{R}_h\mathcal{T}_d$ - see Fig.~\ref{fig1}d. However, $\mathcal{T}_d$ has no effect on $\bf G$ and subsequently no effect on ${\bf M}_n^D$ and ${\bf M}_{\ell,n}$. Therefore the last results apply to symmetries under glide-reflection transformations as well.

To summarize, it has been shown that if the structure possesses inversion symmetry (${\bf p2}$ group), reflection symmetry with respect to the chain axis (${\bf p11m}$ group), or glide-reflection symmetry (${\bf p11g}$ group), an even dispersion is guaranteed even if the particles themselves possess non reciprocal property such as that arises e.g.~by magnetized plasmonic material.

\subsection{The non-uniqueness of the unit cell}
So far we have discussed the symmetries on the unit cell level. However, while the period length, $D$, is unique, the choice of a unit cell is not. The latter always consists of $N$ particles, and there are up to $N$ different choices for the unit cell and even more possible $\bb{M}(\beta)$ (due to the freedom of "numbering" the particles in the same unit cell \cite{Note1}). Let us define the set $\mathscr{M}$ as the set of all possible matrices $\bb{M}(\beta)$ for a certain structure. If the determinant of the matrix $\bb{M}(\beta)$ vanishes for a certain choice of the unit cell, then it must vanish for all $\mathscr{M}$ - since all the matrices in $\mathscr{M}$ represent the wave dynamics of the same structure. Therefore, if the operations done on $\bb{M}(\beta)$ in the previous subsections do not result in the mentioned variations of the same $\bb{M}(\beta)$ but do result in these variations of one of the other matrices in $\mathscr{M}$ then the structure must still possess even dispersion. For example, if the unit cell does not posses an inversion symmetry, but upon inverting the unit cell we obtain a differently structured unit cell, but of the exact same \emph{original, pre-inverted} structure - then the dispersion is rendered even since the substitution $\beta\rightarrow -\beta$ still preserves the zero determinant. A good example for this is the $\bf p11g$ group, displayed in Fig.~\ref{fig1}d. A unit cell does not posses any of the discussed symmetries - invariance under $\mathcal{R}_\pi$ or under $\mathcal{R}_h$. However, under $\mathcal{R}_h$ we obtain a different unit cell of the same structure, and therefore it is reciprocal. This fact helps us finally establish the list of Frieze groups that host even dispersions: ${\bf p11m}$, $\bf p11g$, $\bf p2$, $\bf p2mg$, $\bf p2mm$. Uneven dispersion can only be found in groups $\bf p1$ and $\bf p1m1$.

\subsection{Random period}

Consider a periodic chain whose period is randomly generated (e.g.~randomly locating $N$ particles inside the period). Clearly, for $N=1$ the notion of randomness becomes meaningless. For $N=2$ and if $\bb{\alpha}_1=\bb{\alpha}_2$, one may verify that the period would always be invariant under $\mathcal{R}_\pi$, i.e. it resides in the {\bf p2} group and hence it possesses an even dispersion. Note that this argument holds only for dipole-moment excitation. If higher order excitations become relevant (e.g.~quadrupoles), then the above conclusion may not hold. However, for $N\ge 3$ nearly any random construction would reside in the {\bf p1} group whether the particles are identical or not, yielding uneven dispersion under magnetic bias (see Sec.~\ref{seq:Ex} below).

\subsection{Nonreciprocal sector-way metasurfaces}

A new family of metasurfaces, termed as \emph{metaweaves}, can be obtained by invoking a procedure of ``weaving'' one-way particle chains (one-way ``threads''). The resulting surface possesses a novel ``$\phi$ sector-way'' dynamics: when the surface is excited by a point dipole it allows the propagation of trapped modes only into a cone whose vertex-angle is $\phi$. This structure has been suggested and studied in \cite{Metaweaves}. With the present approach, however, one may design non-reciprocal sector-way metasurfaces without the need to directly ``weave'' one-way threads. Generally, the symmetries of such metasurfaces can be studied in the framework of the wallpaper groups in a manner similar to the study in the previous subsections. However, there are 17 symmetry groups, hence for brevity we use here a rule of thumb (see example below). To achieve sector-way dynamics it is sufficient to create a periodic metasurface whose period does not possess $\mathcal{R}_h,\mathcal{R}_v$, and $\mathcal{R}_\pi$ symmetries (i.e.~the two-dimensional period should not be invariant under these operations).

\subsection{Distinctive features related to modeling of non-reciprocal particle arrays}

\begin{figure}
        \includegraphics[width=7.0cm]{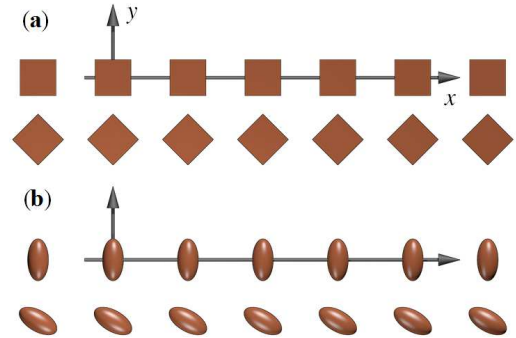}
    \caption{Examples of symmetry considerations in isotropic and non-isotropic particles. (a) A structure made of non-isotropic particles that have isotropic (scalar) polarizability when not magnetized. (b) Non-isotropic particles with non-isotropic polarizabilty.
    }
    \label{fig1ba}
\end{figure}

Consider a particle with no circular symmetry, but with an isotropic (scalar) polarizability when not magnetized. A cube made of isotropic dielectric material is an example for such a particle \cite{Sihvola_Pol}. An example of a structure made of such particles is shown in Fig.~\ref{fig1ba}a.
Formally, this structure resides in the {\bf p1m1} group (or in {\bf p1} group if the particles in the a lower chain are rotated by an angle smaller than $\pi/4$). However, since the polarizabilities are isotropic, this asymmetry will not show up in the particle-dipole model, and the corresponding dispersion is even also under magnetization. It should be emphasized that the theoretical analysis above relate only to the particle's polarizabilities, as seen in e.g.~\Eq{eq8}, \Eq{eq9b}, and \Eq{eq11b}. Hence, the symmetries demonstrated in Fig.~\ref{fig1} should be perceived as symmetries that relate to particle's locations and \emph{polarizabilities} (and not particle geometries). In light of the above, the structure in Fig.~\ref{fig1ba}a would possess uneven dispersion only if the particles are sufficiently large, so that higher order multipoles are significantly excited and are needed to model its dynamics.
To contrast, the structure in Fig.~\ref{fig1ba}b consists of particles that possess anisotropic polarizability. Hence, it resides in {\bf p1} group both in the formal geometrical sense (like the cubes structure), and in the polarizability sense (unlike the cubes structure).

In conclusion of the above discussion, particle structures \emph{may} possess ``hidden'' symmetries that take effect only when the particles are small, and disappear when considering larger particles or continuous medium. Furthermore, the ability to assign ``point-wise'' rotation to a particle whose polarizability is non-isotropic (e.g. ellipsoids), adds degrees of freedom not available in continuous medium or in 1D photonic structure such as the one studied e.g. in \cite{FigotinVitebsky1}.

\section{Examples}\label{seq:Ex}

All the structures in the examples below consist of plasmonic particles, exposed to a uniform transverse magnetization bias $\bb{B}_0=\unit{z}B_0$.
Generally, the single particle polarizability under magnetization can be found, e.g.~in \cite{MazorSteinberg_PlanarOneWay} for metal particles of ellipsoidal shape, and in \cite{Graphene_Magnetized} for engineered Graphene particles using more detailed quantum model based on the Kubo theory \cite{Kubo}. We note that the latter possesses quasi-static resonance very similar to that of the Drude model, with resonant frequency whose role is similar to that of $\omega_p$ in metals, but in the 40-50 THz frequencies. For all particles the off-diagonal terms of $\bb{\alpha}_n$ are imaginary and of opposite signs, i.e.~they satisfy \Eq{eq11b}. The strength of the non-reciprocity in $\bb{\alpha}$ is measured generally by the parameter $\kappa=\abs{\alpha_{xy}/\alpha_{xx}}$. Furthermore, note that $\kappa=\tan\theta$ where $\theta$ is the polarization-rotation angle associated with the non-reciprocity. For metal particles under the Drude model $\kappa$ gets a simple algebraic expression, from which the strength of the magnetization level is easily extracted:
$\kappa=\omega_b/\omega_p$ where $\omega_p$ is the plasma resonance frequency and $\omega_b=eB_0/m_e$ is the cyclotron frequency. For the engineered particles in \cite{Graphene_Magnetized} the dependence of $\kappa$ on $B_0$ is much more complicated - see e.g.~Eqs.~(10)-(11) in the supplementary information of \cite{Graphene_Magnetized}. To get a feeling of the numbers involved, we note that in the examples below we typically have $\kappa\approx 0.0075-0.01$. Under the Drude model and for metals such as Ag, this yields $B_0$ way beyond any practical realization. However, in the particles of \cite{Graphene_Magnetized}, the magnetization strengths required for these values of $\kappa$ is significantly less than 1 Tesla in the relevant frequency regime (see e.g. Figs.~2,3 there).

To keep the algebra simple, the polarizabilities of the magnetized spherical and elliptical particles in all the examples below were calculated according to the formulas provided in \cite{MazorSteinberg_PlanarOneWay}. However, the final algebraic form of $\bb{\alpha}$ in \cite{Graphene_Magnetized} is precisely the same and it needs $B_0\le 1$ Tesla.

Finally, we note that the structures in the present study are inherently non-Bravais lattices. Hence, they possess $2N$ dispersion branches for each plasma resonance; $N$ is the number of particles in a period, and the factor 2 is due to the vector-nature of the particle's dipole moment. To avoid cluttering the dispersion diagrams we show below only the branches where the non-even dispersion was examined and one-way guiding was obtained.

\subsection{p1m1 group}
\subsubsection{By virtue of particle geometry}
\begin{figure}
        \includegraphics[width=7.0cm]{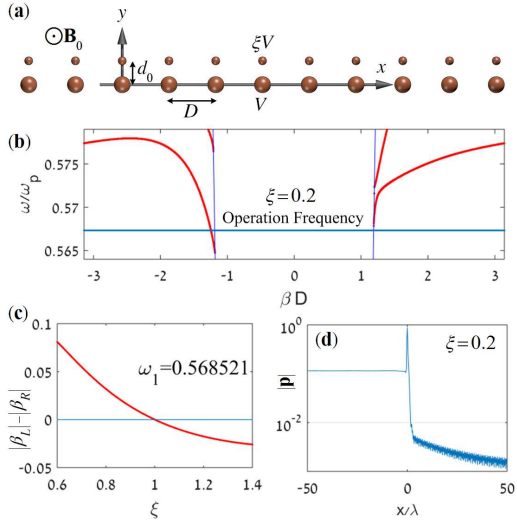}
    \caption{A planar periodic structure of group {\bf p1m1} possessing uneven dispersion under magnetic bias normal to the structure's plane. (a) The structure geometry. (b) The dispersion for $\xi=0.2$. The vertical blue lines are the light cone. (c) A measure of the uneven dispersion. (d) The response to a point dipole located at the center, with the frequency shown by the horizontal blue line in (b).
    }
    \label{fig2}
\end{figure}

Figure \ref{fig2}a shows a structure consisting of two parallel chains of plasmonic particles, with identical period lengths. The chains differ by the corresponding particle's volume: the particles in one chain (lower) are of volume $V$, while those of the other chain (upper) are of volume $\xi V$, $\xi>0$. The corresponding polarizabilities scale accordingly ($\bb{\alpha}$ is proportional to $V$). Clearly, for $\xi\ne 1$ the structure belongs to the $\bf p1m1$ group. The other parameters are $D=\lambda_p/3$, $a=\lambda_p/12$, $d_0=0.8D$, where $\lambda_p$ is the plasma resonance wavelength and $a$ is the particle diameter. The magnetic bias strength corresponds to $\kappa=0.01$ (=$\omega_b/\omega_p$ in Drude metals). The dispersion of plasmonic modes guided by the structure is obtained by looking for the real values of $\beta(\omega)$ that nullify the determinant of \Eq{eq4a}. Figure \ref{fig2}b shows the dispersion for $\xi=0.2$ (the particle's diameter scales as $0.2^{1/3}\approx 0.585$). The uneven nature is evident. To get a feeling of how it depends on the parameter $\xi$, Fig.~\ref{fig2}c compares $\beta(\omega_1)$ on the left side ($\beta_L$) and on the right side ($\beta_R$) of the dispersion, at a frequency $\omega_1$ for which both are real. Only at $\xi=1$ one obtains $\beta_L=\beta_R$. Finally, Fig.~\ref{fig2}d shows the response of the structure when excited by a point dipole located at its center, with the frequency shown by the horizontal line in Fig.~\ref{fig2}b. Note that at this frequency there is only one intersection with a guided mode. This mode has negative $\beta$ and negative group velocity (the red line in Fig.~\ref{fig2}b). There are also two intersections with the light-line cone corresponding to the so called ``light-line modes''. The latter are practically unexcitable in SDC-type structures, whether reciprocal \cite{EnghetaChain,HadadSteinbergPRB}, or non-reciprocal \cite{HadadMazorSteinberg_GreensOneWay}. Hence only $-\unit{x}$ propagating mode can be excited, as seen in the response in Fig.~\ref{fig2}d.

\subsubsection{By virtue of lattice geometry}

\begin{figure}
        \includegraphics[width=7.0cm]{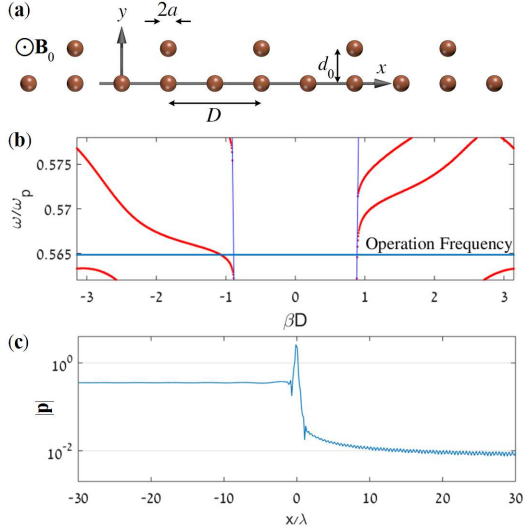}
    \caption{(a) Two parallel chains of identical particles, but with different periods. (b) The structure dispersion (c) The response to a point dipole located at the structure's center, with the frequency shown by the horizontal blue line in (b).
    }
    \label{fig3}
\end{figure}

Figure \ref{fig3}a shows another structure of the {\bf p1m1} group, consisting of two parallel chains of identical particles but with period lengths $D$ and $D/2$. The other parameters are $D=\lambda_p/4$, $a=\lambda_p/24$, $d_0=0.5D$, and $\kappa=0.0075$ (=$\omega_b/\omega_p$ in Drude metals). The uneven dispersion and one-way operation are seen in Figs.~\ref{fig3}b,c.

\subsection{p1 group}

In this group the structure period possesses no symmetry at all. Hence, in principle, it is the easiest to fabricat or synthesize. In fact, nearly \emph{any} period that consists of \emph{randomly located particles}, would result in uneven dispersion under transverse magnetic bias, and consequently in one-way operation. Below we show two examples of ordered and random periods.

\subsubsection{By virtue of lattice geometry}
\begin{figure}
        \includegraphics[width=7cm]{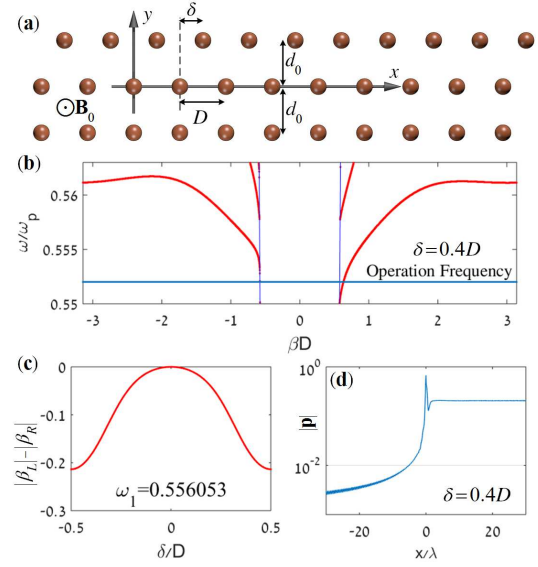}
    \caption{A planar periodic structure of group {\bf p1m1} if $\delta=\pm 0.5D$, and to the {\bf p1} group if $\delta\ne \pm 0.5D, 0$. For $\delta\ne 0$ it possesses uneven dispersion under magnetic bias normal to the structure's plane. (a) The structure geometry. (b) The structure's dispersion for $\delta=0.4d$. (c) A measure of the uneven dispersion. (d) The response to a point dipole located at the structure's center, with the frequency shown by the horizontal blue line in (b).
    }
    \label{fig4}
\end{figure}

The geometry shown in Fig.~\ref{fig4}a consists of three parallel chains of identical particles and equal period length $D$. Two of which are precisely aligned, while the third is shifted along its axis by the distance $\delta$. Note that for $\delta/D=\pm 0.5$ the structure belongs to the {\bf p1m1} group, and for $\delta/D\ne\pm 0.5, 0$ the structure is in the {\bf p1} group. The parameters are: period length $D=\lambda_p/6$, particle radius $a=\lambda_p/24$, distance between the chains $d_0=0.75D$, and magnetic bias level that corresponds to $\kappa=0.01$. Figure \ref{fig4}b shows the dispersion for $\delta=0.4D$. Figure \ref{fig4}c shows $\abs{\beta_L}-\abs{\beta_R}$ at a frequency $\omega_1$ for which guiding exists for both sides ($\omega_1=0.556053\omega_p$), vs.~$\delta/D$ for the range $-0.5\le\delta/D\le 0.5$. Figure~\ref{fig4}d shows the structure's response to a point dipole located at its center.

\begin{figure}
        \includegraphics[width=7.5cm]{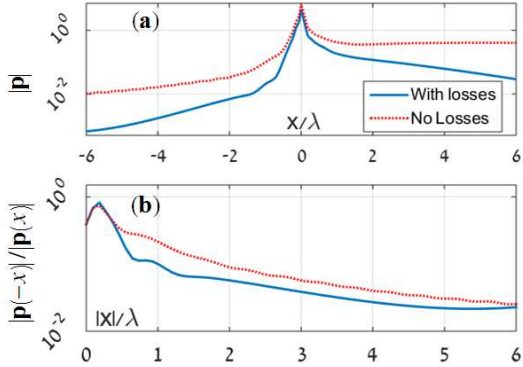}
    \caption{The structure in Fig.~\ref{fig4}, but for Ag particles with loss. (a) Response. (b) The structure \emph{isolation ratio} $\abs{\bb{p}(-x)}/\abs{\bb{p}(x)}$ that provides a measure of the one-way property.
    }
    \label{figLosses}
\end{figure}

To get a feeling of actual array parameters and the effect of loss, we note that when using Ag particles the parameters above correspond to $\lambda_p=150$nm \cite{Metals}, particle diameter is $2a=12$nm, $D=25$nm, and $d_0=19$nm. The loss parameter of Ag is $\tau=0.5\times 10^{-12}$ Sec. Due to the presence of loss $\beta$ becomes complex and dispersion curves such as those in Fig.~\ref{fig4} cannot be drawn. Instead, we use the working point shown in Fig.~\ref{fig4} to simulate the structure response. The results are shown in Fig.~\ref{figLosses}. It is seen that while the signal itself decays as it propagates along the structure, the one way property is still profound and survives the loss, as evident from the the \emph{isolation ratio}. Achieving the required $\kappa$ under the Drude model for Ag requires magnetization levels that are far beyond practical values. However, we note that essentially the same physical dimensions would apply for structures made of the particles studied in \cite{Graphene_Magnetized} ($2a\approx 10$ to 20 nm and central frequency in the range of 40-50 THz), with $B_0$ in the order of 1 Tesla or less.

\subsubsection{By virtue of particle geometry and randomness}

In all the examples so far, our structures consist of at least two parallel periodic chains. Can one obtain uneven dispersion using a single chain? It is clear that as long as the particles themselves are spherical, invariance of a single perfectly linear chain under $\mathcal{R}_h$ cannot be broken. Hence such structures belong to the {\bf p11m} group (at least) and their dispersion would always be even.

However, if the particles are \emph{not} precise spheres (e.g.~ellipsoids), one may assign the property of \emph{geometrical rotation} to each particle. This property is not invariant under reflection, hence it opens the way to achieve uneven dispersion using a single chain. This approach of using \emph{chiral chains} to achieve one-way plasmonic guiding has been suggested and studied in
\cite{HadadSteinberg_SpiralChain,MazorSteinberg_PlanarOneWay,HadadMazorSteinberg_GreensOneWay}. Since it incorporates geometrical rotation and Faraday rotation simultaneously, it is termed as the \emph{two-type rotation} principle. We refer the reader to \cite{MazorSteinberg_PlanarOneWay,HadadMazorSteinberg_GreensOneWay} for examples of planar chains. Hence, in the context of the present study, the two-type rotation principle can be viewed as a method to generate {\bf p1} or {\bf p1m1} structures systematically, using only a single chain. The studies in \cite{MazorSteinberg_PlanarOneWay,HadadMazorSteinberg_GreensOneWay} dealt with structures where the rotation of the $n$-th ellipsoidal particle is given by $\theta_n=n\Delta\theta$. If $\Delta\theta/\pi$ is rational, this would result in a perfectly periodic chain.

\begin{figure}
        \includegraphics[width=7cm]{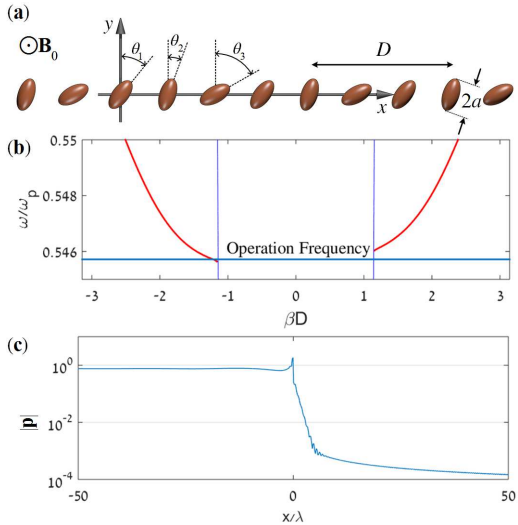}
    \caption{An example of a periodic chain of randomly rotated ellipsoidal particles. (a) A schematic view of the structure. (b) The dispersion for $(\theta_1,\theta_2,\theta_3)=(0.329,0.794,0.468)\pi$. (c) Response to a dipole excitation.
    }
    \label{fig5}
\end{figure}

The generalization  of the previous studies, as suggested by the present work, is not only of formal nature. It may have also important practical implications. In fact, almost every \emph{random} generation of a period, would end up in {\bf p1} group and may provide one-way guiding. We demonstrate this important finding within the family of the two-type rotation structures. Figure \ref{fig5}a shows, schematically, a chain of rotated prolate ellipsoids whose period $D$ consists of three particles. The ellipsoids rotation angles are \emph{chosen at random}. In the specific example here we have $(\theta_1,\theta_2,\theta_3)=(0.329,0.794,0.468)\pi$. Other parameters are: $D=\lambda_p/3$,
$a=\lambda_p/27$ (prolate's major axis), $b=0.9a$ (prolate's minor axis), and $\kappa=0.0075$. Figure \ref{fig5}b shows the corresponding dispersion, and Fig.~\ref{fig5}c the chain response to a point dipole excitation at its center. One way behavior is evident, with isolation of nearly $10^{-4}$ over distances of 5-6 wavelengths.

\subsection{Nonreciprocal sector-way metasurfaces}

\begin{figure}
        \includegraphics[width=7cm]{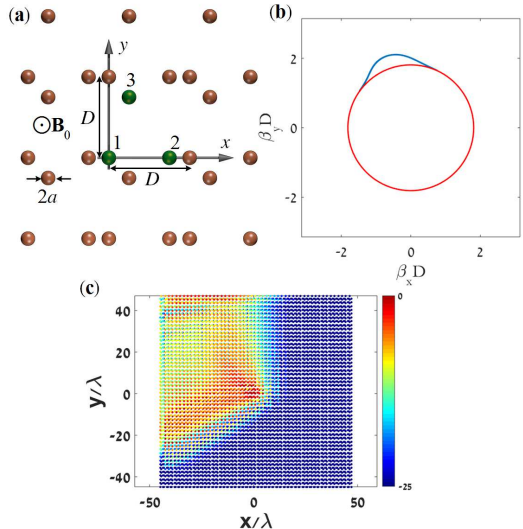}
    \caption{The metaweave. (a) Lattice geometry, possessing no symmetry under $\mathcal{R}_h,\mathcal{R}_v$, and $\mathcal{R}_\pi$. Under normal magnetic bias this structure possesses sector-way guiding. (b) A dispersion contour. The red circle in the center is the light-cone. (c) Response to a point-dipole excitation.
    }
    \label{fig6}
\end{figure}

An example of a sector-way metasurface, or meta-weave, is shown in Fig.~\ref{fig6}a. It is a generalization of the structures suggested in \cite{Metaweaves}.
The structure period in both directions is $D$, and it consists of only three spherical particles located at $\rv_1=(0,0)$, $\rv_2=(0.75,0)D$, and $\rv_3=(0.25,0.75)D$. Clearly, this period does not possess $\mathcal{R}_h,\mathcal{R}_v$, and $\mathcal{R}_\pi$ symmetries. We have calculated the dispersion surfaces and dipole response of the structure with the following parameters: $D=\lambda_p/2$, $a=\lambda_p/24$, and $\kappa=0.005$ (the weakest magnetic bias in our examples). Figure \ref{fig6}b shows the dispersion contour for the frequency $\omega=0.5762249\omega_p$. Note the asymmetric shape: it is not invariant under the map $\bb{\beta}\mapsto-\bb{\beta}$. The circle in the center is the light-cone. Figure \ref{fig6}c shows the response to an excitation by point dipole located at the center, with the frequency corresponding to Fig.~\ref{fig6}b. Sector-way response is evident.

\section{Conclusions}

General symmetry considerations were used to predict the conditions under which one may obtain uneven dispersion curves an optical isolation in particle arrays. The idiosyncracies of dipole-dipole interactions, related to asymmetry in particle arrays were discussed. It has been shown that under magnetic bias, structures belonging to the {\bf p1} and {\bf p1m1} Frieze groups shown in Fig.~\ref{fig1}, possess this dispersion asymmetry. The uneven dispersion can then be utilized for optical isolation. The single chain structure for one-way guiding, based on the two-type rotation principle, is shown to be a special case where one excludes the necessary symmetries by assigning geometrical rotation to each lattice point. Hence, it provides the narrowest optical isolator. It is further shown that almost any random periods with more than two particles would result in uneven dispersion.

The results of this work provide a set of rules that can be useful for designing optica isolators, and as building blocks for optical circulators. Since one-way guiding inherently supresses propagation in the ``wrong'' direction, applications that require minimization of back-reflections may benefit. These applications can be, for example, new ways to feed electromagnetic components as antennas and resonators in a perfectly matched fashion \cite{Engheta_MatchedLoad, HadadSteinberg_Antenna}.

\appendix
\section{Matrices symmetry properties}\label{AppSymm}

The free-space electric field $\bb{E}(\rv)$ due to a point dipole $\bb{p}$ at $\rv'$ in is given by
\begin{eqnarray}\label{eqAppG1}
\!\!\! {\bf E}(\rv) &=& \epsilon_0^{-1}
G(R)\left\{k^2(\unit{n}\times\bb{p})\times\unit{n}\vphantom{\frac{1}{R^2}}\right. \nonumber\\
&+&\left.\left[3 \unit{n} (\unit{n}\cdot\bb{p})-\bb{p}\right]\left(\frac{1}{R^2}-\frac{ik}{R}\right)\right\}
\end{eqnarray}
where $G(R)=(4\pi R)^{-1}e^{ikR}$, $R=\abs{\rv-\rv'}$, and $\unit{n}=(\rv-\rv')/R$. From the above, the free-space Green's function matrix used in e.g.~\Eq{eq2} can be written as
\begin{eqnarray}\label{eqAppG2}
{\bf G}(\rv,\rv') &=& \epsilon_0^{-1}G(R)
\left[k^2\left({\bf I}-\bf{U}\right)\vphantom{\frac{1}{R^2}}\right. \nonumber\\
 &+& \left. \left( 3{\bf U}-{\bf I}\right)\left(\frac{1}{R^2}-\frac{ik}{R}\right)\right]
\end{eqnarray}
where ${\bf I}$ is the identity matrix, and ${\bf U}=\unit{n}^T\unit{n}$. ${\bf G}(\rv,\rv')$ satisfies the following symmetry relations
\begin{equation}\label{eqAppSymm1}
{\bf G}(\rv,\rv')={\bf G}^T(\rv,\rv')={\bf G}(\rv',\rv)
\end{equation}
that are a manifestation of the free-space reciprocity and symmetry. If $\rv,\rv'$ are restricted to the $z=0$ plane, then $\bf G$ is reduced to a symmetric $2\times 2$ matrix
\begin{subequations}
\begin{equation}\label{eqAppG3a}
{\bf G}(\rv,\rv')=\left(\begin{array}{ll}
g_{xx} & g_{xy}\\
g_{xy} & g_{yy}
\end{array}\right),
\end{equation}
satisfying
\begin{eqnarray}
{\bf G}(\mathcal{R}_h\rv,\mathcal{R}_h\rv') &=& \left(\begin{array}{cc}
g_{xx} & -g_{xy}\\
-g_{xy} & g_{yy}
\end{array}\right)\nonumber\\
& &\nonumber\\
 &=& {\bf I^h}_2 {\bf G}(\rv,\rv') {\bf I^h}_2 \label{eqAppG3b}
\end{eqnarray}
where ${\bf I^h}_2=\mbox{diag}(-1,1)$. The last equality above deals with reflection about the $x$ axis. Similar relation holds for the operation $\mathcal{R}_v$. Furthermore, inversion leaves $\bf G$ unchanged
\begin{equation}\label{eqAppG3c}
{\bf G}(\mathcal{R}_\pi\rv,\mathcal{R}_\pi\rv')={\bf G}(\rv,\rv')
\end{equation}
\end{subequations}

We turn to discuss the matrices ${\bf M}_n^D$ and ${\bf M}_{n,\ell}$, defined in \Eqs{eq4b}{eq4c}. Due to the periodicity of our structure we have, $\forall\, m,n,\ell$ :
\begin{equation}\label{eqAppSymm2}
\rv_{0,n}-\rv_{m,n}=\rv_{-m,\ell}-\rv_{0,\ell}=-\unit{x}mD
\end{equation}
The first symmetries of ${\bf M}_n^D$ and ${\bf M}_{n,\ell}$ emerge only from \Eq{eqAppSymm1} and from the strip periodicity. For ${\bf M}_n^D(\beta)$ we have,
\begin{eqnarray}\label{eqAppSymm3}
{\bf M}_n^D(-\beta) &=& \sumex{m}{}{0}{\bf G}(\rv_{0,n},\rv_{m,n})e^{-im\beta D}\nonumber\\
                    &=&
\sumex{m}{}{0}{\bf G}(\rv_{0,n},\rv_{-m,n})e^{im\beta D}
\end{eqnarray}
We use now \Eq{eqAppSymm2} with $\ell=n$, and \Eq{eqAppSymm1} in \Eq{eqAppSymm3}. The result is \Eq{eq5}.

Regarding ${\bf M}_{n,\ell}(-\beta)$,
\begin{eqnarray}\label{eqAppSymm4}
{\bf M}_{n,\ell}(-\beta) &=& \sum_m{\bf G}(\rv_{0,n},\rv_{m,\ell})e^{-im\beta D}\nonumber\\
                    &=&
\sum_m{\bf G}(\rv_{0,n},\rv_{-m,\ell})e^{im\beta D}
\end{eqnarray}
however, from \Eq{eqAppSymm2}
$\rv_{0,n}-\rv_{-m,\ell}=\rv_{m,n}-\rv_{0,\ell}$. By using this result and \Eq{eqAppSymm1} in \Eq{eqAppSymm4}, we obtain \Eq{eq6}.

{\bf \emph{Inversion and reflection symmetries}}
If the structure possesses more symmetries than that of a mere periodicity, their footprint appear as more symmetries of ${\bf M}_n^D$ and ${\bf M}_{n,\ell}$. To explore them let us number the points within a period from left to right and from bottom to top. Then, the set of points in a period $\rv_{0,1},\rv_{0,2},\ldots,\rv_{0,N}$ possesses an inversion symmetry if $\forall n=1,\ldots,N$,
\begin{equation}\label{eqAppSymm8}
\mathcal{R}_\pi\rv_{0,n}\equiv-(x_{0,n},y_{0,n})=\rv_{0, N+1-n}.
\end{equation}
Using this result together with \Eq{eqAppSymm2}, \Eq{eqAppG3c}, and \Eqs{eq4b}{eq4c}, we obtain for inversion symmetric set of points ($n,\ell=1,\ldots,N$)
\begin{subequations}
\begin{eqnarray}
{\bf M}^D_{N+1-n} &=&  {\bf M}^D_n                  \label{eqAppSymm9a}\\
{\bf M}_{\ell,n}  &=&  {\bf M}_{N+1-n, N+1-\ell}.   \label{eqAppSymm9b}
\end{eqnarray}
\end{subequations}
Likewise, if the set of points possesses horizontal reflection symmetry
\begin{equation}\label{eqAppSymm10}
\mathcal{R}_h\rv_{0,n}\equiv(x_{0,n},-y_{0,n})=\rv_{0, N+1-n}
\end{equation}
then, using \Eq{eqAppG3b} in the definitions of ${\bf M}_n^D$ and ${\bf M}_{n,\ell}$ we obtain
\begin{subequations}
\begin{eqnarray}
{\bf I^h}_2 {\bf M}_n^D {\bf I^h}_2 &=& {\bf M}_{N+1-n}^D \label{eqAppSymm11a}\\
{\bf I^h}_2 {\bf M}_{n,\ell} {\bf I^h}_2 &=& {\bf M}_{N+1-n,N+1-\ell} \label{eqAppSymm11b}
\end{eqnarray}
\end{subequations}

\section{Inversion and reflection operations}\label{AppInv}

Let $\bb{R}$ presents a set of $N$ points in the plane; $\bb{R}=(\rv_1,\rv_2,\ldots\, \rv_N)=(x_1,y_1,x_2,y_2,\ldots, x_n,y_N)$. Assume that the points are numbered, e.g.~from left to right and from bottom to top. Then, up to an arbitrary linear shift of the entire set, an inversion $\mathcal{R}_\pi$ of the set of points can be characterized by the matrix ${\bf I^i}_{2N}$
\begin{equation}\label{eqAppInv1}
\mathcal{R}_\pi\bb{R}={\bf I^i}_{2N}\bb{R}^T
\end{equation}
where the non-zero entries of ${\bf I^i}_{2N}$ occupy only the secondary block-diagonal elements, and are given by $N$ conventional $2\times 2$ identity matrices ${\bf I}_2$ [${\bf I}_2=\mbox{diag}(1,1)$],
\begin{equation}\label{eqAppInv2}
{\bf I^i}_{2N}=-\left(
\begin{array}{lllll}
0 & 0 & \ldots & 0 & {\bf I}_2\\
0 & 0 & \ldots & {\bf I}_2 & 0\\
\vdots & & & 0 & 0\\
{\bf I}_2& 0 & \ldots & 0& 0
\end{array}\right)
\end{equation}
We note that $\abs{{\bf I^i}_{2N}}=1$, hence operation by ${\bf I^i}$ is determinant preserving.
Likewise, we note that reflection about the $x$ axis--horizontal reflection $\mathcal{R}_h$ is characterized by the matrix ${\bf I^h}_{2N}$
\begin{equation}\label{eqAppInv3}
\mathcal{R}_h\bb{R}={\bf I^h}_{2N}\bb{R}^T
\end{equation}
where ${\bf I^h}_{2N}=\mbox{diag}({\bf I^h}_2,{\bf I^h}_2,\ldots,{\bf I^h}_2)$,
and ${\bf I^h}_2=\mbox{diag}(-1,1)$.

\end{document}